\documentclass[12pt]{article}

\usepackage{scicite}
\usepackage{graphicx}
\usepackage{graphics}
\usepackage{epsfig}
\usepackage{color}
\usepackage{times}

\newenvironment{sciabstract}{%
\begin{quote} \bf}
{\end{quote}}

\newcounter{lastnote}

\title{Response to comment by Tin-Lun Ho on ``Itinerant Ferromagnetism in a Strongly Interacting Fermi Gas of Ultracold Atoms", \textit{Science} \textbf{325}, 1521 (2009)}

\author
{Gyu-Boong Jo$^{1}$, Ye-Ryoung Lee$^{1}$, Jae-Hoon Choi$^{1}$, Caleb A. Christensen$^{1}$, \\
Tony H. Kim$^{1}$, Joseph H. Thywissen$^{2}$, David E. Pritchard$^{1}$, and Wolfgang Ketterle$^{1}$\\
\\
\normalsize{$^{1}$MIT-Harvard Center for Ultracold Atoms, Research
Laboratory of Electronics,}\\
\normalsize{ Department of Physics, Massachusetts
Institute of Technology, Cambridge, MA 02139, USA}\\
\normalsize{$^{2}$Department of Physics, University of Toronto, Toronto, Ontario M5S1A7, Canada}\\
\\
\normalsize{}
}

\date{}

\begin{document}


\baselineskip15pt


\maketitle
% define commands for international characters7
\catcode`\ä = \active \catcode`\ö = \active \catcode`\ü = \active
\catcode`\Ä = \active \catcode`\Ö = \active \catcode`\Ü = \active
\catcode`\ß = \active \catcode`\é = \active \catcode`\è = \active
\catcode`\ë = \active \catcode`\ô = \active \catcode`\ê = \active
\catcode`\ø = \active \catcode`\ò = \active \catcode`\í = \active
\catcode`\Ó = \active \catcode`\ú = \active \catcode`\á = \active
\catcode`\ã = \active
\defä{\"a} \defö{\"o} \defü{\"u} \defÄ{\"A} \defÖ{\"O} \defÜ{\"U} \defß{\ss} \defé{\'{e}}
\defè{\`{e}} \defë{\"{e}} \defô{\^{o}} \defê{\^{e}} \defø{\o} \defò{\`{o}} \defí{\'{i}}
\defÓ{\'{O}} \defú{\'{u}} \defá{\'{a}} \defã{\~{a}}
\newcommand{\li}{$^6$Li}
\newcommand{\na}{$^{23}$Na}
\newcommand{\cs}{$^{133}$Cs}
\newcommand{\kk}{$^{40}$K}
\newcommand{\rb}{$^{87}$Rb}
\newcommand{\vect}[1]{\mathbf #1}
\newcommand{\g}{g^{(2)}}
\newcommand{\one}{$\left|1\right>$}
\newcommand{\two}{$\left|2\right>$}
\newcommand{\V}{V_{12}}
\newcommand{\kfa}{\frac{1}{k_F a}}

\begin{sciabstract}

Ho claims in his comment that our experiment is direct evidence that itinerant ferromagnetism does not exist in ultracold Fermi gases.  This claim is incorrect and based on an invalid estimate of relaxation times and an erroneous interpretation of the detectability of ferromagnetic domains.  We point out that the experimental evidence is consistent with the existence of ferromagnetism, but further experiments are needed to distinguish a ferromagnetic ground state from a non-magnetic ground state with ferromagnetic correlations.

\end{sciabstract}

In our recent paper~\cite{Jo09}, we showed for a Fermi gas of lithium-6 atoms that the lifetime, kinetic energy, and cloud size vary non-monotonously for increasing repulsive interactions, and that this behavior is consistent with predictions of a phase transition to a ferromagnetic state based on mean-field models.  However, we were not able to observe ferromagnetic domains due to finite imaging resolution and line of sight integration, which suggests that the size of domains were smaller than 2$\mu$m.

We explicitly state in our paper that all our measurements are sensitive only to local spin polarization and are independent of domain structure.  This implies that further experimental evidences are required to distinguish between equilibrium domains and short-rage fluctuating domains.  In our conclusion, we explicitly point out that our interpretation in terms of a phase transition to itinerant ferromagnetism is based on the qualitative agreement with the prediction of simple models~\cite{meanfield}.  We also stated that strong interactions and correlations, for which no detailed theoretical treatment exists, could possibly modify our findings.

The possible importance of correlations is reiterated by Ho.  However, no theoretical treatment is provided in his comment. Instead, he refers to theoretical studies in lattices, but it is not clear how they can be applied to the continuum case studied in our paper. Recent work~\cite{Zhai09_correlated} shows, within a phenomenological model, that correlations can lead to similar experimental signatures as we have observed. However, the model does not quantitatively agree with our data, and has some qualitative discrepancies as well; for example the extrema of lifetime, kinetic energy and cloud size don't occur at the same value of the parameter $k_F a$, where $k_F$ is the Fermi momentum and $a$ the s-wave scattering length, in contrast to our observations. It would be interesting to see if further development of the alternative theories could lead to quantitative agreement with experiments. Note that a ferromagnetic phase transition has been predicted by theories including mean-field and correlations in second order~\cite{Duine05,conduit09}, but it remains to be seen if correlations are adequately treated.

Ho argues that our non-observation of spin domains clearly shows their absence and therefore the existence of a non-magnetic state.  He states that even if the formation of domains favors small sizes, there should be occasionally a domain which is large enough to be detected.  However, Ho makes no predictions about the statistics of occurrence of large domains, and how they could be detected in the presence of statistical and systematic noise sources.  We note that our non-observation of domains was based on visual inspections of several images which didn't show any discernable textures.  We presented these results in our paper using weak language (for example, ``a signal-to-noise ratio ... suggests" and ``we suspect..." ) and gave estimates without any error bars.  Ho's suggestion of looking for the rare event of a large domain conflicts with interference fringes, speckle and other imaging artifacts which, at some level, are present in all experimental images. To exclude the existence of such domains, one needs a prediction about their probability of occurrence, and a careful analysis of all experimental limitations.

Recent work~\cite{demler09}, posted prior to Ho's comment, makes predictions about the size distribution of domains and their growth rate after a rapid quench across a critical value of $k_F a$.  These authors conclude that the expected pattern size of $\sim2{k_F}^{-1}$ is much smaller than the experimental imaging resolution and provide a theoretical explanation for the non-observation of domains.

Finally, Ho claims that large domains should form on a time scale $\hbar/E_F$, where $\hbar$ is Planck's constant divided by $2\pi$ and $E_F$ the Fermi energy, which is fast compared to the hold time in our experiment.  This estimate is incorrect. It should apply only to the local response, i.e. screening of interactions and local correlations, but not to the formation of domains, which should show a slowing down near the quantum critical point. The time of domain formation must depend on their size $l$, so there is a second dimensionless parameter $k_F l$  in the problem.  This is directly confirmed in the calculations of Ref.~\cite{demler09} which predict that the time scale for domain formation strongly depends on the domain size and how far the system is quenched beyond the critical point.

In conclusion, we strongly disagree with Ho that our experiment has shown that Fermi gases with strong repulsive interactions are non-magnetic. Ho's claim ignores the dependence of the time scale for domain formation on their size. So far, the experimental evidence is consistent with a phase transition to a ferromagnetic state, but it cannot rule out a non-magnetic state with strong ferromagnetic correlations, partly due to the fact that no detailed theoretical predictions exist for such a state.

We thank Eugene Demler for valuable discussions.

\bibliography{FMreference_Jo}

\bibliographystyle{Science}

\end{document}